\documentclass[conference,10pt]{IEEEtran}
\IEEEoverridecommandlockouts

\usepackage{epsfig,rotating,setspace,latexsym,amsmath,epsf,amssymb,amsfonts,bm,theorem,cite,algorithm,graphicx,epsf,authblk,epstopdf,color,algpseudocode,bbm}

\newtheorem{corollary}{Corollary}

\newtheorem{lemma}{Lemma}
\newenvironment{Proof}[1]{\medskip\par\noindent{\bf Proof:\,}\,#1}{{\mbox{\,$\blacksquare$}\par}}

\allowdisplaybreaks

\begin{document}

\title{Timely Cloud Computing: Preemption and Waiting}

\author[1]{Ahmed Arafa}
\author[2]{Roy D. Yates}
\author[1]{H. Vincent Poor}
\affil[1]{\normalsize Electrical Engineering Department, Princeton University}
\affil[2]{\normalsize Department of Electrical and Computer Engineering, Rutgers University}

\maketitle

\begin{abstract}
The notion of timely status updating is investigated in the context of cloud computing. Measurements of a time-varying process of interest are acquired by a sensor node, and uploaded to a cloud server to undergo some required computations. These computations consume random amounts of service time that are independent and identically distributed across different uploads. After the computations are done, the results are delivered to a monitor, constituting an {\it update}. The goal is to keep the monitor continuously fed with fresh updates over time, which is assessed by an {\it age-of-information} (AoI) metric. A scheduler is employed to optimize the measurement acquisition times. Following an update, an idle waiting period may be imposed by the scheduler before acquiring a new measurement. The scheduler also has the capability to {\it preempt} a measurement in progress if its service time grows above a certain {\it cutoff} time, and upload a fresher measurement in its place. Focusing on stationary deterministic policies, in which waiting times are deterministic functions of the instantaneous AoI and the cutoff time is fixed for all uploads, it is shown that the optimal waiting policy that minimizes the long term average AoI has a threshold structure, in which a new measurement is uploaded following an update only if the AoI grows above a certain threshold that is a function of the service time distribution and the cutoff time. The optimal cutoff is then found for standard and shifted exponential service times. While it has been previously reported that waiting before updating can be beneficial for AoI, it is shown in this work that preemption of {\it late} updates can be even more beneficial. 
\end{abstract}

\section{Introduction}

We consider the problem of timely computing. The setting is motivated by some applications in which monitoring a time-varying process of interest can be computationally demanding. Hence, instead of extracting useful information from local data measurements acquired by sensor nodes, measurements are uploaded to a cloud server that can handle heavy-duty computation tasks, and send them back in the form of updates. Computation times, however, are random, and the process may have already changed by the time they are done. We therefore investigate the benefits of preempting an upload in progress and replacing it by a new, {\it fresher}, one. Such freshness/timeliness is assessed by the {\it age-of-information} (AoI), defined as the time elapsed since the latest received update.

\begin{figure}[t]
\center
\includegraphics[scale=.9]{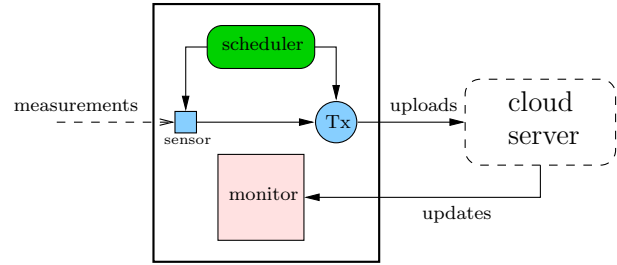}
\caption{A scheduler decides on when to acquire new measurements by the sensor and send them to the cloud server by the transmitter. The server updates the system's monitor after it completes the required computations.}
\label{fig_sys_mod_cloud}
\vspace{-.1in}
\end{figure}

Lots of work pertaining to AoI minimization have appeared in the recent literature, with frameworks that include queuing, optimization and scheduling, energy harvesting, remote estimation, and coding, see, e.g., \cite{yates_age_1, ephremides_age_random, sun-age-mdp, talak-age-interference, zhou-age-iot, zhang-arafa-aoi-pricing-wiopt, melih-aoi-soft, batu-aoi-multihop, jing-age-online, arafa-age-online-finite, elif-age-online-threshold, ornee-aoi-estimation-ou, yates-age-erase-code, arafa-aoi-coding}. Of particular relevance to our work are those in \cite{kaul-aoi-lcfs, ephremides_age_management, chen-age-error, yates-age-mltpl-src, najm-age-multistream, soysal-aoi-gg11, farazi-aoi-eh-preempt, kavitha-aoi-preempt, wang-aoi-preempt}, which show that the notion of preemption of updates in service and replacing them by new ones is viable for AoI minimization in various settings, which is mainly owing to the nature of AoI that promotes sending fresh updates. This is discussed in a queuing framework in \cite{kaul-aoi-lcfs, ephremides_age_management, chen-age-error}, and more recently in \cite{yates-age-mltpl-src} that also extends to the case of multiple sources. Different from \cite{kaul-aoi-lcfs, ephremides_age_management, chen-age-error, yates-age-mltpl-src} that focus on exponential service times, the work in \cite{najm-age-multistream} considers general service time distributions for multiple Poisson sources with preemption. Preemption for general arrival and service time distributions, for a single source, has been recently studied in \cite{soysal-aoi-gg11}. Reference \cite{farazi-aoi-eh-preempt} characterizes settings for which preemption is age-minimal, subject to energy harvesting constraints with Poisson arrivals (of both energy and updates) and exponential service times. The studies in \cite{kavitha-aoi-preempt, wang-aoi-preempt} investigate a similar tradeoff, under different system models, namely, that while preemption lets the system work with the freshest information, it leads to restarting service from the beginning. Thus, a decision has to be made on whether to drop the newly arriving updates or switch to them via preemption. Recently, in the context of computing, AoI analysis has been carried out through various tandem queuing models in \cite{xu-aoi-tandem-preprocess, kuang-aoi-tandem-mec, gong-aoi-mec, zou-aoi-tandem-preprocess}, and through a task-specific age metric in \cite{song-aoi-task-mec}. The notion of sending timely measurements to the cloud has been discussed in the context of gaming in \cite{yates-aoi-cloud-gaming}.

In this paper we investigate the tradeoff in \cite{kavitha-aoi-preempt, wang-aoi-preempt} in a cloud computing setting. Different from \cite{kavitha-aoi-preempt, wang-aoi-preempt}, however, we consider a {\it generate-at-will} model, in which measurement times are controlled by a scheduler. Each measurement is uploaded to a cloud server to undergo some computations before being sent back as an update. {\it The scheduler has the ability to preempt a measurement in service if its computation time is larger than a certain cutoff time and upload a fresher one instead.} After an update is eventually received, the acquisition of a new measurement may be scheduled after an idle waiting period. We note that due to preemption, the optimal waiting policy derived in \cite{sun-age-mdp} does not apply in our setting.

%
%
%
%
%

Focusing on stationary deterministic policies, in which cutoff times are constant and waiting times are function of the instantaneous AoI, we show that optimal waiting has a {\it threshold} structure. Specifically, a new measurement is uploaded, following an update, only if the AoI grows above a certain threshold that is a function of the cutoff time and the service time distribution. Such function is given in {\it closed-form}. We also provide a necessary and sufficient condition on the optimality of zero-wait policies, in which a new measurement is uploaded just-in-time as an update is received. When zero-wait is not optimal, we provide a a relatively simple method of evaluating the long term average AoI through a bisection search. We then discuss the evaluation of the optimal cutoff time explicitly under exponential service times. Finally, we compare the proposed preemption and waiting scheme to three baselines: no preemption and zero-waiting; no preemption and optimal waiting, the scheme proposed in \cite{sun-age-mdp}; and optimal preemption and zero-waiting. While it is demonstrated that our proposed scheme perfoms best, our results also show that, depending on the system parameters, the optimal preemption and zero-waiting policy can actually beat the no preemption and optimal waiting policy. This sheds light on the fact that, in some situations, working with fresh measurements provides the highest gains in terms of AoI.

\section{System Model and Problem Formulation}

We consider a system comprised of a sensor, a scheduler, a transmitter, a cloud computing server and a monitor. The overall goal is to keep the system's monitor continuously fed with {\it fresh} status updates pertaining to a physical phenomenon of interest. Such updates, however, require some heavy-duty computations on the raw data measurements acquired by the sensor that need to be carried out by the cloud computing server. Therefore, in order for a status update to reach the monitor, the following series of events need to occur. First, the scheduler decides on when to acquire a new data measurement by the sensor, and send (upload) it to the server by the transmitter. The server then undertakes the computations and feeds back the end result to the monitor in the form of an {\it update}. Hence, the goal is to design a scheduling policy such that these updates reach the monitor in a timely manner. A depiction of the system model considered is shown in Fig.~\ref{fig_sys_mod_cloud}. Hereafter, we will refer to data sent to the server by uploads, and data received from the server by updates.

Uploads are time-stamped so that when updates eventually reach the monitor, the system knows when their corresponding measurements were acquired. We use an AoI metric to assess the timeliness of updates at the monitor. This is defined as
\begin{align}
a(t)=t-u(t),
\end{align}
where $u(t)$ is the time stamp of the latest update that has reached the monitor. 

To minimize the AoI, measurements are uploaded to the cloud server right away after being acquired by the sensor. We assume that upload transmission times are negligible. However, each measurement consumes a computational time at the cloud server denoted as the service time. Service times of different measurements are independent and identically distributed (i.i.d.) according to the distribution of a random variable $X$. Depending on the application or the task being considered, the server may incur a constant delay before the actual computation starts. Let us denote such time by $c\in\mathbb{R}_+$, which is the largest constant such that
\begin{align} \label{eq_srv_lb}
X\geq c\quad \text{a.s.}
\end{align}
This is without loss of generality since $X\geq0$ a.s.\footnote{One might consider the constant $c$ a necessary overhead to initiate service at the cloud server for each measurement.} Motivated by freshness, the scheduler is capable of {\it preempting} the current upload in service if its service time surpasses a certain {\it cutoff} time. Thus, an update will reach the monitor only if its service ends within the cutoff time. Following a preemption, a new measurement is taken and uploaded immediately. Following an update delivery, however, the scheduler may {\it wait} for some idle time before uploading a new measurement.


\begin{figure}[t]
\center
\includegraphics[scale=.8]{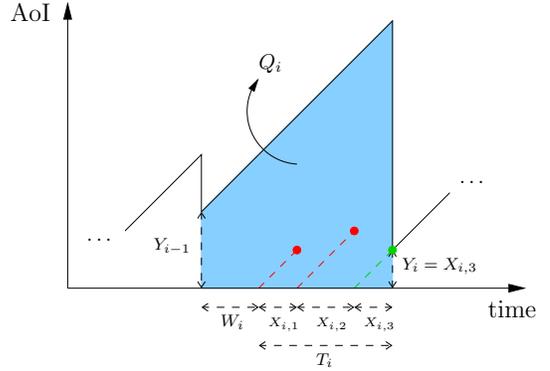}
\caption{AoI evolution example in the $i$th epoch. Red lines denote preemptions and the green line denotes completed service. In this example $N_i=3$.}
\label{fig_age_xmpl_cloud}
\vspace{-.2in}
\end{figure}

We denote by the $i$th epoch the time in between the reception of the $(i-1)$th and the $i$th updates. The $i$th epoch starts with age $Y_{i-1}$ and ends with age $Y_i$.\footnote{We assume that the first epoch starts with some given age $Y_0$ at time $0$.} A waiting period of $W_i$ time units follows after the epoch starts, through which the server is idle. After that, the first measurement in the epoch is acquired and uploaded to the server. Note that, depending on the preemption policy, there can be multiple uploads during a single epoch. We denote by $N_i$ the number of uploads during the $i$th epoch, with $N_i\geq1$, and $N_i-1$ denoting the number of preemptions. Let $X_{i,j}$ denote the service time of the $j$th upload during the $i$th epoch, $1\leq j\leq N_i$. Note that $X_{i,j}$'s are i.i.d $\sim X$. Now observe that only the $X_{i,N_i}$ service time period concludes with an update sent back to the monitor, and all other service time periods end with a preemption. Therefore, the $i$th epoch ends with age
\begin{align}
Y_i=X_{i,N_i}.
\end{align}
We denote by $T_i$ the server's busy period in the $i$th epoch, defined as
\begin{align}
T_i\triangleq X_{i,1}+X_{i,2}+\dots+X_{i,N_i}.
\end{align}
Lastly, let $L_i$ denote the $i$th epoch length given by
\begin{align}
L_i=W_i+T_i.
\end{align}
In Fig.~\ref{fig_age_xmpl_cloud}, we show an example sample path of how the AoI may evolve during the $i$th epoch. From the figure, the area under the AoI curve during the $i$th epoch, $Q_i$, is given by
\begin{align}
Q_i=Y_{i-1}L_i+\frac{1}{2}L_i^2.
\end{align}

We are interested in minimizing the long term average area under the AoI curve. It is clear that such quantity depends on the choices of the cutoff and waiting times that the scheduler makes. Let $\gamma_{i,j}$ denote the cutoff time after which the $j$th upload in the $i$th epoch is preempted. In other words, given $\gamma_{i,j}$, the scheduler preempts the $j$th upload in the $i$th epoch if $X_{i,j}$ grows above $\gamma_{i,j}$ time units. Clearly,
\begin{align}
\gamma_{i,j}\geq c
\end{align}
must hold $\forall i,j$ in view of (\ref{eq_srv_lb}). The set $\{\gamma_{i,j}\}$ now constitutes a {\it cutoff policy}, while the set $\{W_i\}$ denotes a {\it waiting policy}. Let $\pi$ denote a scheduling policy $\{W_i,\gamma_{i,j}\}$. The goal is to solve the following problem:
\begin{align} \label{opt_main}
\min_{\pi}\limsup_{n\rightarrow\infty}\frac{\sum_{i=1}^n\mathbb{E}\left[Q_i\right]}{\sum_{i=1}^n\mathbb{E}\left[L_i\right]},
\end{align}
where $\mathbb{E}\left[\cdot\right]$ is the expectation operator.

\section{Stationary Deterministic Policies}

Observe that the optimal policy $\pi^*$ that solves problem (\ref{opt_main}) may be such that the waiting and cutoff times of the $i$th epoch depend on the history of events, e.g., AoI evolution, number of preemptions, service time realizations, before, and during, the $i$th epoch. To alleviate the difficulty of tracking all such history, and motivated by the fact that service times are i.i.d., we focus on policies that are characterized by the following two main features: $1)$ the waiting time $W_i$ in the $i$th epoch is given by a {\it deterministic} function of the starting AoI $Y_{i-1}$,
\begin{align}
W_i\triangleq w\left(Y_{i-1}\right),
\end{align}
for some function $w:\mathbb{R}_+\rightarrow\mathbb{R}_+$; and $2)$ the cutoff times $\{\gamma_{i,j}\}$ in the $i$th epoch are given by {\it deterministic} functions of the instantaneous AoI,
\begin{align}
\gamma_{i,j}\triangleq \gamma_j\left(a_{i,j}\right),
\end{align}
for some function $\gamma_j:\mathbb{R}_+\rightarrow [c,\infty]$, $\forall j$, with $a_{i,j}$ denoting the AoI just before the $j$th upload occurs in the $i$th epoch.


Let $\Pi_s$ denote the set of policies that abide by the above structure. Note that any $\pi\in\Pi_s$ induces {\it stationary} distributions $Q_i\sim Q$ and $L_i\sim L$ for all epochs. Therefore, under $\Pi_s$, problem (\ref{opt_main}) reduces to
\begin{align} \label{opt_s}
\min_{\pi\in\Pi_s}\frac{\mathbb{E}\left[Q\right]}{\mathbb{E}\left[L\right]}.
\end{align}
Problem (\ref{opt_s}) is an optimization problem over a single epoch. In the sequel, we drop the index $i$ for convenience. We now have the following lemma:

\begin{lemma} \label{thm_gamma_const}
In the optimal solution of problem (\ref{opt_s}), all cutoff functions are equivalent. That is,
\begin{align}
\gamma_j\left(a_j\right)\equiv\gamma\left(a_j\right),\quad\forall j,
\end{align}
for some $\gamma:\mathbb{R}_+\rightarrow [c,\infty]$. 
\end{lemma}

\begin{Proof}
Let the optimal cutoff function $\gamma_1(\cdot)$ be given. Note that the system is idle before the first upload. Thus, $\gamma_1(a_1)$ represents the optimal cutoff time for the AoI to evolve starting from an idle state at age $a_1$. Now assume that the first upload is preempted after $\gamma_1(a_1)$, whence the age becomes $a_2=a_1+\gamma_1(a_1)$. Observe that the system becomes {\it instantly} idle right before the second upload occurs. Since service times are i.i.d., therefore $\gamma_2(a_2)$ should also represent the optimal cutoff time for the AoI to evolve starting from an idle state at age $a_2$. This shows that $\gamma_2(a_2)=\gamma_1(a_2)$ must hold, otherwise $\gamma_1(\cdot)$ would not be optimal. Similar arguments follow for $\gamma_j(\cdot)$, $j\geq3$. Therefore, all cutoff functions are equivalent.
\end{Proof}

In the sequel, we further focus on the case in which the cutoff function $\gamma(\cdot)$ is a constant. That is, with a slight abuse of notation,
\begin{align}
\gamma\left(a_j\right)=\gamma,\quad\forall j,
\end{align}
for some $\gamma\geq c$. We call this the {\it $\gamma$-cutoff policy.} Considering such policy is motivated by the fact that service times are i.i.d.; it also sets a fixed maximum value of $\gamma$ on the starting AoI of each epoch.

Now let the following quantities be (re)defined for the epoch in consideration: $\overline{Y}$ is the starting AoI; $W=w\left(\overline{Y}\right)$ is the waiting time after it starts; $T$ is the server's busy period; $X_j$ is the $j$th upload service time; $N$ is the total number of uploads; and $\underline{Y}$ is the ending AoI. Observe that under a $\gamma$-cutoff policy, given $N=n$, $X_1=\dots=X_{n-1}=\gamma$ and $X_n=\underline{Y}\leq\gamma$ a.s. Also observe that the function $w(\cdot)$ is now restricted to the domain $[0,\gamma]$, and that $\overline{Y}$ and $\underline{Y}$ are i.i.d $\sim Y$. To evaluate the distribution of the age $Y$, let us define $p\triangleq\mathbb{P}\left(X\leq\gamma\right)$, where $\mathbb{P}\left(\cdot\right)$ is the probability measure. Therefore the probability distribution function (PDF) of $Y$ is given by 
\begin{align}
f_Y(y)=\begin{cases}\frac{f_X(y)}{p},\quad&c\leq y\leq\gamma \\
0,\quad&\text{otherwise}\end{cases},
\end{align}
where $f_X(\cdot)$ denotes the PDF of the service time $X$.\footnote{We focus on continuous random variables, and assume that $\gamma$ and the distribution of $X$ are such that $p>0$.}

We note that problem (\ref{opt_s}) is structurally different from the setting considered in \cite{sun-age-mdp}. There, an epoch could only consist of one packet in service until it finishes, and hence the AoI at the end of the epoch relates to that packet's acquisition time. In our setting, owing to the preemption capability, there can be multiple uploads in a single epoch, and hence the AoI at the end of the epoch does not necessarily relate to the first upload time. The optimal waiting policy derived in \cite{sun-age-mdp}, therefore, does not apply in our setting.

Solving problem (\ref{opt_s}) is tantamount to characterizing the optimal waiting function $w^*\left(\cdot\right)$ and the optimal cutoff time $\gamma^*$. In the next sections, we do so sequentially as follows: we first characterize $w^*\left(\cdot\right)$ for a fixed value of $\gamma$, and then we determine $\gamma^*$ for specific service time distributions.

\section{Threshold Waiting Policy} \label{sec_wait}

In this section, we evaluate the optimal waiting policy for fixed cutoff time $\gamma$. The main result is that the optimal waiting policy exhibits a threshold structure, in which a new upload occurs only if the AoI grows above a certain threshold that depends on the service time distribution and the fixed cutoff time. Toward showing that, we need to evaluate some expressions first. We start with
\begin{align}
\mathbb{P}\left(N=n\right)=(1-p)^{n-1}p,\quad n\geq1,
\end{align}
i.e., $N$ is a geometric random variable with parameter $p$. It is useful to note that $\mathbb{E}\left[N\right]=\frac{1}{p}$ and $\mathbb{E}\left[N^2\right]=\frac{2-p}{p^2}$. Using iterated expectations, we now have
\begin{align}
\mathbb{E}\left[T\right]=&\sum_{n=1}^\infty\mathbb{P}\left(N=n\right)\mathbb{E}\left[X_1+X_2+\dots+X_n\right] \nonumber \\
=&\sum_{n=1}^\infty\mathbb{P}\left(N=n\right)\left((n-1)\gamma+\mathbb{E}\left[\underline{Y}\right]\right) \nonumber \\
=&\left(\frac{1}{p}-1\right)\gamma+\mathbb{E}\left[\underline{Y}\right]. \label{eq_exp_x}
\end{align}
Thus, the expected epoch length is given by
\begin{align} \label{eq_exp_L}
\mathbb{E}\left[L\right]=\mathbb{E}\left[w\left(\overline{Y}\right)\right]+\mathbb{E}\left[T\right],
\end{align}
with $\mathbb{E}\left[T\right]$ given by (\ref{eq_exp_x}). Proceeding similarly, we have
\begin{align}
\mathbb{E}&\left[T^2\right]=\sum_{n=1}^\infty\mathbb{P}\left(N=n\right)\mathbb{E}\left[\left(X_1+X_2+\dots+X_n\right)^2\right] \nonumber \\
=&\sum_{n=1}^\infty\mathbb{P}\left(N=n\right)\left((n-1)^2\gamma^2+2(n-1)\gamma\mathbb{E}\left[\underline{Y}\right]+\mathbb{E}\left[\underline{Y}^2\right]\right) \nonumber \\
=&\left(\frac{2-p}{p^2}-\frac{2}{p}+1\right)\gamma^2+2\left(\frac{1}{p}-1\right)\gamma\mathbb{E}\left[\underline{Y}\right] +\mathbb{E}\left[\underline{Y}^2\right]. \label{eq_exp_x2}
\end{align}
We now have
\begin{align}
\mathbb{E}\left[Q\right]=&\mathbb{E}\left[\overline{Y}\left(w\left(\overline{Y}\right)+T\right)\right]+\frac{1}{2}\mathbb{E}\left[\left(w\left(\overline{Y}\right)+T\right)^2\right] \nonumber \\
=&\mathbb{E}\left[\overline{Y}w\left(\overline{Y}\right)\right]+\mathbb{E}\left[\overline{Y}\right]\mathbb{E}\left[T\right]+\frac{1}{2}\mathbb{E}\left[w^2\left(\overline{Y}\right)\right] \nonumber \\
&+\mathbb{E}\left[w\left(\overline{Y}\right)\right]\mathbb{E}\left[T\right]+\frac{1}{2}\mathbb{E}\left[T^2\right], \label{eq_exp_Q}
\end{align}
with $\mathbb{E}\left[T\right]$ and $\mathbb{E}\left[T^2\right]$ given by (\ref{eq_exp_x}) and (\ref{eq_exp_x2}), respectively, and the second equality follows by independence of $\overline{Y}$ and $T$.

To find the optimal $w^*(\cdot)$, we need to solve the following functional optimization problem:
\begin{align} \label{opt_fxd_g}
\min_{w(\cdot)}&\quad\frac{\mathbb{E}\left[Q\right]}{\mathbb{E}\left[L\right]} \nonumber \\
\mbox{s.t.}&\quad w(t)\geq0,\quad c\leq t\leq\gamma.
\end{align}
To solve the above problem, we follow Dinkelbach's approach \cite{dinkelbach-fractional-prog} and introduce the following auxiliary problem for some fixed parameter $\lambda\geq0$:
\begin{align} \label{opt_fxd_g_aux}
g(\lambda)\triangleq\min_{w(\cdot)}&\quad\mathbb{E}\left[Q\right]-\lambda\mathbb{E}\left[L\right] \nonumber \\
\mbox{s.t.}&\quad w(t)\geq0,\quad c\leq t\leq\gamma.
\end{align}
One can show that $g(\lambda)$ is decreasing in $\lambda$, and that the optimal solution of problem (\ref{opt_fxd_g}) is given by $\lambda^*$ that solves $g(\lambda^*)=0$ \cite{dinkelbach-fractional-prog}. By monotonicity of $g(\cdot)$, $\lambda^*$ can be found by, e.g., a bisection search. Focusing on problem (\ref{opt_fxd_g_aux}), we introduce the following Lagrangian \cite{boyd}:
\begin{align}
\mathcal{L}=\mathbb{E}\left[Q\right]-\lambda\mathbb{E}\left[L\right]-\int_c^\gamma w(\tau)\eta(\tau)d\tau,
\end{align}
where $\eta(\cdot)$ is a Lagrange multiplier. Substituting (\ref{eq_exp_L}) and (\ref{eq_exp_Q}) above, and after some rearrangements we get
\begin{align} \label{eq_lagrangian}
\mathcal{L}=&\int_c^\gamma\left(\left(\tau+\mathbb{E}\left[T\right]-\lambda\right)w(\tau)+\frac{1}{2}w^2(\tau)\right)f_Y(\tau)d\tau \nonumber \\
&+\mathbb{E}\left[\overline{Y}\right]\mathbb{E}\left[T\right]+\frac{1}{2}\mathbb{E}\left[T^2\right]-\lambda\mathbb{E}\left[T\right]-\int_c^\gamma w(\tau)\eta(\tau)d\tau.
\end{align}

Now taking the (functional) derivative of $\mathcal{L}$ with respect to $w(t)$, $c\leq t\leq\gamma$, and equating to $0$ we have
\begin{align}
\left(t+\mathbb{E}\left[T\right]-\lambda+w^*(t)\right)f_Y(t)-\eta(t)=0.
\end{align}
Rearranging the above, we get that
\begin{align} \label{eq_w_eta}
w^*(t)=\lambda-\mathbb{E}\left[T\right]-t+\frac{\eta(t)}{f_Y(t)}.
\end{align}
We note that there are different methods through which one can conclude that the optimal waiting policy satisfies (\ref{eq_w_eta}). These are discussed in Appendix~\ref{apndx_w_eta} for completeness. Now using complementary slackness \cite{boyd}, (\ref{eq_w_eta}) further gives
\begin{align} \label{eq_w}
w^*(t)=\left[\lambda-\mathbb{E}\left[T\right]-t\right]^+,\quad c\leq t\leq\gamma,
\end{align}
where $\left[\cdot\right]^+\triangleq\max(\cdot,0)$. This makes the AoI right after the waiting period, when the first measurement in the epoch gets uploaded, equal to
\begin{align}
t+w^*(t)=\max\{t,\lambda-\mathbb{E}\left[T\right]\},
\end{align}
which comes directly from the fact that $w^*(t)>0$ if and only if (iff) $\lambda-\mathbb{E}\left[T\right]>t$. Observe that the value of $t$, the realization of $\overline{Y}$, represents the AoI at the beginning of the epoch. Hence, one could interpret the optimal waiting policy as a {\it threshold} policy, in which the first measurement in the epoch gets uploaded only if the AoI grows above $\lambda-\mathbb{E}\left[T\right]$.

To have an operational significance, however, the threshold $\lambda-\mathbb{E}\left[T\right]$ must be positive. The next lemma verifies that this is indeed the case. The proof is in Appendix~\ref{apndx_thm_lmda_lb}.

\begin{lemma} \label{thm_lmda_lb}
The optimal solution of problem (\ref{opt_fxd_g}), $\lambda^*$, satisfies $\lambda^*>\mathbb{E}\left[T\right]$.
\end{lemma}

Observe that while Lemma~\ref{thm_lmda_lb} shows that the threshold is positive, a zero-wait policy can still be optimal if the threshold's value is no larger than $c$. The next lemma quantifies this relationship. The proof is in Appendix~\ref{apndx_thm_zero_wait_c}.

\begin{lemma} \label{thm_zero_wait_c}
A zero-wait policy, in which $w^*(t)=0$, $\forall t\in\left[c,\gamma\right]$, is optimal for problem (\ref{opt_fxd_g}) iff
\begin{align} \label{eq_zero_wait_c}
\frac{\frac{1}{2}\left(\frac{1}{p}-1\right)\gamma^2+\frac{1}{2}\mathbb{E}\left[Y^2\right]}{\left(\frac{1}{p}-1\right)\gamma+\mathbb{E}\left[Y\right]}\leq c.
\end{align}
\end{lemma}

The optimal AoI under a zero-wait policy is directly given by substituting $w^*(t)=0$, $\forall t$ in (\ref{eq_exp_L}) and (\ref{eq_exp_Q}) to get
\begin{align}
\lambda^*_{zw}&=\frac{\mathbb{E}\left[Q\right]}{\mathbb{E}\left[L\right]} \nonumber \\
&=\frac{\mathbb{E}\left[\overline{Y}\right]\mathbb{E}\left[T\right]+\frac{1}{2}\mathbb{E}\left[T^2\right]}{\mathbb{E}\left[T\right]} \nonumber \\
&=\mathbb{E}\left[Y\right]+\frac{\frac{1}{2}\mathbb{E}\left[T^2\right]}{\mathbb{E}\left[T\right]}, \label{eq_aoi_zw}
\end{align}
where the subscript $zw$ stands for zero-wait.

Now that we established a necessary and sufficient condition for the optimality of a zero-wait policy in Lemma~\ref{thm_zero_wait_c}, we proceed by investigating the case in which the inequality condition in (\ref{eq_zero_wait_c}) does {\it not} hold. First, an immediate corollary follows in this case.

\begin{corollary} \label{thm_lmda_lb_c}
The optimal solution of problem (\ref{opt_fxd_g}), $\lambda^*$, satisfies $\lambda^*>\mathbb{E}\left[T\right]+c$ iff (\ref{eq_zero_wait_c}) does not hold.
\end{corollary}


\begin{figure}[t]
\center
\includegraphics[scale=1]{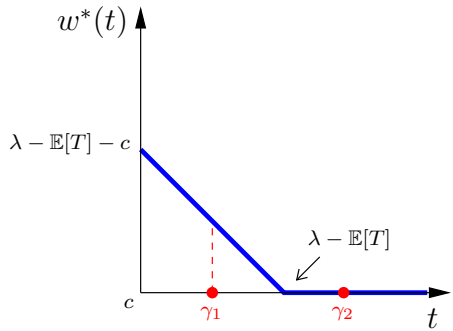}
\caption{The optimal waiting policy versus time. We show two example choices of $\gamma$ in red. For $\gamma_1$, we always wait before uploading a new measurement following an update, while for $\gamma_2$ it depends on the value of $t$. Lemma~\ref{thm_lmda_ub} shows that the situation of $\gamma_1$ cannot be optimal.}
\label{fig_wait_policy}
\end{figure}

Now observe that for $\gamma<\lambda^*-\mathbb{E}\left[T\right]$, one would {\it always} wait before uploading a new measurement following an update, and that for $\gamma\geq\lambda^*-\mathbb{E}\left[T\right]$, it depends on the realization of $\overline{Y}$ (the value of $t$) as indicated in (\ref{eq_w}). We illustrate this behavior in Fig.~\ref{fig_wait_policy}, and settle this issue in the next lemma by showing that the situation of $\gamma_1$ in Fig.~\ref{fig_wait_policy} {\it cannot} be optimal.\footnote{We note that Fig.~\ref{fig_wait_policy} is only explanatory and that in reality the choice of $\gamma$ also affects the values of $\lambda^*$ and $\mathbb{E}\left[T\right]$.} We note that the result of the lemma holds regardless of whether (\ref{eq_zero_wait_c}) holds or not. The proof is in Appendix~\ref{apndx_thm_lmda_ub}.

\begin{lemma} \label{thm_lmda_ub}
The optimal solution of problem (\ref{opt_fxd_g}), $\lambda^*$, satisfies $\gamma\geq\lambda^*-\mathbb{E}\left[T\right]$.
\end{lemma}

In summary, to find the optimal AoI for fixed $\gamma$ one should start by examining (\ref{eq_zero_wait_c}). If it holds, then $\lambda^*=\lambda^*_{zw}$ in (\ref{eq_aoi_zw}). Else, using Corollary~\ref{thm_lmda_lb_c} and Lemma~\ref{thm_lmda_ub}, one has the following bounds on the optimal AoI:
\begin{align} \label{eq_lmda_bounds}
\mathbb{E}\left[T\right]+c<\lambda^*\leq\mathbb{E}\left[T\right]+\gamma,
\end{align}
which facilitates evaluating $\lambda^*$ that solves $g(\lambda^*)=0$ using a bisection search in the interval $\left(\mathbb{E}\left[T\right]+c,\mathbb{E}\left[T\right]+\gamma\right]$.

Now it remains to choose the best $\gamma$ that minimizes $\lambda^*$. We discuss this in the next section.

\section{Optimal $\gamma$-Cutoff Policy}

It is not direct to get a closed-form expression of the optimal $\lambda^*$ in terms of $\gamma$ for general service time distributions. In fact, even for specific distributions this can also be a difficult task. In this section, our goal is to provide some insight on the service time distribution can affect the choice of the optimal cutoff $\gamma^*$. To avoid confusion, let the optimal AoI as a function of the cutoff value, derived in Section~\ref{sec_wait}, be denoted by $\lambda^*\left(\gamma\right)$, and define $\lambda^{**}\triangleq\lambda^*\left(\gamma^*\right)$. Our approach will be as follows: we will first fix $\gamma\geq c$ and evaluate $\lambda^*\left(\gamma\right)$ as discussed  toward the end of Section~\ref{sec_wait}; and then we will evaluate $\gamma^*$ that minimizes $\lambda^*\left(\gamma\right)$, i.e., achieves $\lambda^{**}$, numerically.

We will consider an exponential service time distribution with $c=0$ along with its shifted version with $c>0$. Clearly, the zero-wait policy is not optimal for distributions with $c=0$, as inferred from the inequality (\ref{eq_zero_wait_c}). In this case, $\lambda^*\left(\gamma\right)$ can be evaluated by a bisection search using the bounds in (\ref{eq_lmda_bounds}). On the other hand, for $c>0$, $\lambda^*\left(\gamma\right)$ is given in closed-form by $\lambda^*_{zw}$ in (\ref{eq_aoi_zw}) for values of $\gamma$ that satisfy (\ref{eq_zero_wait_c}), and is evaluated by a bisection search using the bounds in (\ref{eq_lmda_bounds}) otherwise. As we will see, in some situations evaluating $\gamma^*$ will be a direct consequence of evaluating the bounds in (\ref{eq_lmda_bounds}).

\subsection{Standard Exponential}

Let $X\sim\exp(1)$. Since $c=0$, we aim at evaluating the bounds in (\ref{eq_lmda_bounds}). Toward that, one can directly compute the following quantities:
\begin{align}
p=&1-e^{-\gamma}, \\
\mathbb{E}\left[Y\right]=&\frac{1-\left(1+\gamma\right)e^{-\gamma}}{1-e^{-\gamma}}.
\end{align}
This directly gives $\mathbb{E}\left[T\right]=1$, and hence
\begin{align}
1<\lambda^*\left(\gamma\right)\leq1+\gamma,
\end{align}
upon which one can see that $\gamma^*$ is infinitesimal. As mentioned before, this is one instance where evaluating the bounds in (\ref{eq_lmda_bounds}) directly gives $\gamma^*$. Therefore, in this case, $\lambda^{**}$ can be made arbitrarily close to $1$ by choosing $\gamma^*$ arbitrarily close to $0$.

\subsection{Shifted Exponential}

We now focus on the shifted version of the above, in which 
\begin{align}
f_X(x)=e^{-(x-c)}, \quad x\geq c>0.
\end{align}
Based on this, for $\gamma\geq c$, one can directly compute
\begin{align}
p=&1-e^{-\left(\gamma-c\right)}, \\
\mathbb{E}\left[Y\right]=&\frac{1+c-\left(1+\gamma\right)e^{-\left(\gamma-c\right)}}{1-e^{-\left(\gamma-c\right)}}, \\
\mathbb{E}\left[Y^2\right]=&\frac{2+2c+c^2-\left(2+2\gamma+\gamma^2\right)e^{-\left(\gamma-c\right)}}{1-e^{-\left(\gamma-c\right)}}.
\end{align}
Upon substituting all the above in (\ref{eq_zero_wait_c}) and simplifying, we get that the zero-wait policy is optimal iff
\begin{align}
1-\frac{1}{2}c^2\leq\left(1+\gamma-c\right)e^{-\left(\gamma-c\right)}.
\end{align}
Observe that the above is satisfied for all values of $\gamma\geq c$ if $c\geq\sqrt{2}$. Next, note that $\left(1+\gamma-c\right)e^{-\left(\gamma-c\right)}$ is decreasing in $\gamma$, and has a maximum value of $1$ when $\gamma=c$. This shows that there exists a unique $\bar{\gamma}(c)>c$ that satisfies the above inequality with equality if $c<\sqrt{2}$. Thus, the inequality is satisfied for $c<\sqrt{2}$ iff $\gamma\leq\bar{\gamma}(c)$. Based on the above, the zero-wait policy is optimal in the following cases: 1) $c\geq\sqrt{2}$, or 2) $c<\sqrt{2}$ and $\gamma\leq\bar{\gamma}(c)$. On the other hand the zero-wait policy is not optimal if $c<\sqrt{2}$ and $\gamma>\bar{\gamma}(c)$.

\begin{figure}[t]
\center
\includegraphics[scale=.45]{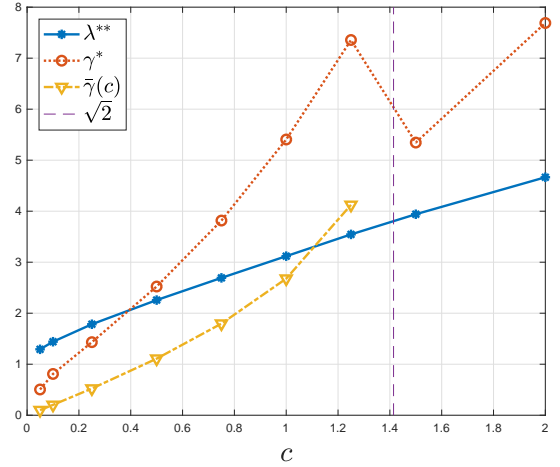}
\caption{Optimal AoI and cutoff values versus $c$ for exponential service times. The vertical line denotes the critical value of $c=\sqrt{2}$, after which the zero-wait policy is optimal.}
\label{fig_exponential_ex}
\end{figure}

In Fig.~\ref{fig_exponential_ex}, we plot the optimal cutoff $\gamma^*$ and the corresponding AoI $\lambda^*$ versus $c$. We also show $\bar{\gamma}(c)$ on the figure to indicate whether zero-wait is optimal for $c<\sqrt{2}$. We see from the figure that the zero-wait policy is {\it not} optimal for all values of $c<\sqrt{2}$ since $\gamma^*>\bar{\gamma}(c)$; it is only optimal for $c\geq\sqrt{2}$. Note that $\bar{\gamma}(c)$ is not defined (and not needed) for $c\geq\sqrt{2}$, and is therefore not shown on the figure.

\begin{figure}[t]
\center
\includegraphics[scale=.45]{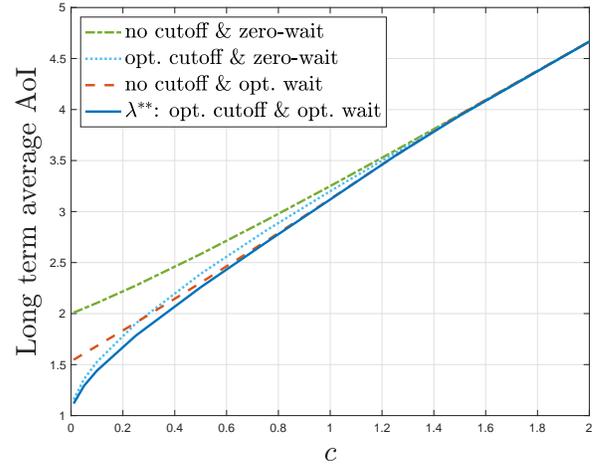}
\caption{Comparing the optimal policy to other bench marks versus $c$ for exponential service times.}
\label{fig_exponential_ex_cmpr}
\end{figure}

In Fig.~\ref{fig_exponential_ex_cmpr}, we compare the optimal policy derived in this paper to other benchmarks. The first is the vanilla version of status updating, denoted {\it no cutoff \& zero-wait}, in which an upload is never preempted, and a new upload takes place once an update is received. The second is also a zero-wait policy yet with optimizing the cutoff value, denoted {\it optimal cutoff \& zero-wait}. The third is that of \cite{sun-age-mdp}, denoted {\it no cutoff \& optimal wait}, in which the waiting time is optimized and uploads are never preempted. We see that our policy beats all benchmarks, especially for small values of $c$. Another interesting note is that for for $c\lessapprox0.25$, optimizing the cutoff turns out to be better, age-wise, than optimizing the waiting time. Indeed, the {\it optimal cutoff \& zero-wait} policy beats the {\it no cutoff \& optimal wait} policy of \cite{sun-age-mdp} for $c\lessapprox0.25$.

\section{Conclusion}

A cloud computing status updating system has been considered, in which computations are carried out on raw data measurements uploaded to a cloud server, and then returned in the form of updates to a monitor. Using an AoI metric, it has been shown that preemption of late updates, whose service times exceed a certain cutoff time, and replacing them by fresher measurements can enhance the overall AoI. Further, it has been shown that it is optimal to upload a new measurement to the server following an update only if the AoI grows above a certain threshold. Implications of such preemption and waiting policies have been discussed for exponential service time distributions, along with comparison with other benchmarks.

\appendix

\subsection{Proof of Lemma~\ref{thm_lmda_lb}} \label{apndx_thm_lmda_lb}

We show this by contradiction. Assume that $\lambda^*\leq\mathbb{E}\left[T\right]$. Then this would necessarily mean that $w^*(t)=0$, $\forall t$, and hence (cf. (\ref{eq_aoi_zw}))
\begin{align}
\lambda^*&=\frac{\mathbb{E}\left[\overline{Y}\right]\mathbb{E}\left[T\right]+\frac{1}{2}\mathbb{E}\left[T^2\right]}{\mathbb{E}\left[T\right]} \nonumber \\
&=\mathbb{E}\left[T\right]-\left(\frac{1}{p}-1\right)\gamma+\frac{\frac{1}{2}\mathbb{E}\left[T^2\right]}{\mathbb{E}\left[T\right]}, \label{eq_lmda_pf_1}
\end{align}
where (\ref{eq_lmda_pf_1}) follows by (\ref{eq_exp_x}). Now for $\lambda^*$ to be no larger than $\mathbb{E}\left[T\right]$, it must hold that
\begin{align} \label{eq_pf_thm_lmda_lb_1}
\frac{\frac{1}{2}\mathbb{E}\left[T^2\right]}{\mathbb{E}\left[T\right]}\leq\left(\frac{1}{p}-1\right)\gamma.
\end{align}
Using (\ref{eq_exp_x}) and (\ref{eq_exp_x2}), the above is tantamount to having
\begin{align} \label{eq_pf_thm_lmda_lb_2}
&\frac{1}{2}\left(\frac{2-p}{p^2}-\frac{2}{p}+1\right)\gamma^2+\left(\frac{1}{p}-1\right)\gamma\mathbb{E}\left[\underline{Y}\right]+\frac{1}{2}\mathbb{E}\left[\underline{Y}^2\right] \nonumber \\
&\hspace{.7in}\leq\left(\frac{1}{p}-1\right)^2\gamma^2+\left(\frac{1}{p}-1\right)\gamma\mathbb{E}\left[\underline{Y}\right],
\end{align}
which, upon some direct algebraic rearrangements, is equivalent to having
\begin{align} \label{eq_pf_thm_lmda_lb_3}
\frac{1}{2}\left(\frac{1}{p}-1\right)\gamma^2+\frac{1}{2}\mathbb{E}\left[\underline{Y}^2\right]\leq0,
\end{align}
which is a clear contradiction.

\subsection{Proof of Lemma~\ref{thm_zero_wait_c}} \label{apndx_thm_zero_wait_c}

In view of (\ref{eq_w}), a zero-wait policy is optimal iff $\lambda^*\leq\mathbb{E}\left[T\right]+c$. Proceeding as in Appendix~\ref{apndx_thm_lmda_lb}, this is tantamount to adding $c$ to the right hand side (RHS) of (\ref{eq_pf_thm_lmda_lb_1}), or equivalently adding $c\mathbb{E}\left[T\right]$ to the RHSs of (\ref{eq_pf_thm_lmda_lb_2}) and (\ref{eq_pf_thm_lmda_lb_3}). Thus, a zero-wait policy is optimal iff
\begin{align}
\frac{\frac{1}{2}\left(\frac{1}{p}-1\right)\gamma^2+\frac{1}{2}\mathbb{E}\left[\underline{Y}^2\right]}{\mathbb{E}\left[T\right]}\leq c.
\end{align}
Substituting (\ref{eq_exp_x}) above directly gives (\ref{eq_zero_wait_c}).

\subsection{Proof of Lemma~\ref{thm_lmda_ub}} \label{apndx_thm_lmda_ub}

First, if (\ref{eq_zero_wait_c}) holds, then by Corollary~\ref{thm_lmda_lb_c} $\lambda^*\leq\mathbb{E}\left[T\right]+c\leq\mathbb{E}\left[T\right]+\gamma$.

We now show the result of the lemma when (\ref{eq_zero_wait_c}) does not hold. We show this by contradiction. Assume that $\gamma<\lambda^*-\mathbb{E}\left[T\right]$. Under that assumption, it holds by (\ref{eq_w}) that
\begin{align}
w^*(t)=\lambda-\mathbb{E}\left[T\right]-t,\quad c\leq t\leq\gamma,
\end{align}
i.e., $w^*(t)>0$, $\forall t\in[c,\gamma]$. Therefore,
\begin{align}
\mathbb{E}\left[w\left(\overline{Y}\right)\right]=&\int_c^\gamma \left(\lambda-\mathbb{E}\left[T\right]-\tau\right)f_Y(\tau)d\tau \nonumber \\
=&\lambda-\mathbb{E}\left[T\right]-\mathbb{E}\left[\overline{Y}\right]. \label{eq_exp_w_pf_ub}
\end{align}

Our goal now is to evaluate the value of $\lambda^*$ by solving $g(\lambda^*)=0$, and show that it cannot be larger than $\mathbb{E}\left[T\right]+\gamma$, thereby reaching a contradiction. Toward that, we start by using the above to evaluate 
\begin{align}
\mathbb{E}\left[L\right]=\mathbb{E}\left[w\left(\overline{Y}\right)\right]+\mathbb{E}\left[T\right]=\lambda-\mathbb{E}\left[\overline{Y}\right]. \label{eq_thm_exp_L}
\end{align}
Since $\mathbb{E}\left[L\right]\geq0$, it must hold that the optimal $\lambda^*$ satisfies
\begin{align}
\lambda^*\geq\mathbb{E}\left[\overline{Y}\right]. \label{eq_thm_lmda_lb}
\end{align}
This simple observation will prove to be useful later on.

Next, we have
\begin{align}
\mathbb{E}\left[\overline{Y}w\left(\overline{Y}\right)\right]=&\int_c^\gamma\tau\left(\lambda-\mathbb{E}\left[T\right]-\tau\right)\frac{f_Y(\tau)}{p}d\tau \nonumber \\
=&\left(\lambda-\mathbb{E}\left[T\right]\right)\mathbb{E}\left[\overline{Y}\right]-\mathbb{E}\left[\overline{Y}^2\right], \label{eq_exp_yw_pf_ub}
\end{align}
and
\begin{align}
\mathbb{E}\left[w^2\left(\overline{Y}\right)\right]=&\int_c^\gamma\left(\lambda-\mathbb{E}\left[T\right]-\tau\right)^2\frac{f_Y(\tau)}{p}d\tau \nonumber \\
=&\left(\lambda-\mathbb{E}\left[T\right]\right)^2-2\left(\lambda-\mathbb{E}\left[T\right]\right)\mathbb{E}\left[\overline{Y}\right]+\mathbb{E}\left[\overline{Y}^2\right]. \label{eq_exp_w2_pf_ub}
\end{align}
Substituting (\ref{eq_exp_w_pf_ub}), (\ref{eq_exp_yw_pf_ub}) and (\ref{eq_exp_w2_pf_ub}) in (\ref{eq_exp_Q}) we get
\begin{align}
\mathbb{E}\left[Q\right]=&\left(\lambda-\mathbb{E}\left[T\right]\right)\mathbb{E}\left[\overline{Y}\right]-\mathbb{E}\left[\overline{Y}^2\right]+\mathbb{E}\left[\overline{Y}\right]\mathbb{E}\left[T\right] \nonumber \\
&+\frac{1}{2}\left(\lambda-\mathbb{E}\left[T\right]\right)^2-\left(\lambda-\mathbb{E}\left[T\right]\right)\mathbb{E}\left[\overline{Y}\right]+\frac{1}{2}\mathbb{E}\left[\overline{Y}^2\right] \nonumber \\
&+\left(\lambda-\mathbb{E}\left[T\right]-\mathbb{E}\left[\overline{Y}\right]\right)\mathbb{E}\left[T\right]+\frac{1}{2}\mathbb{E}\left[T^2\right] \nonumber \\
=&-\frac{1}{2}\mathbb{E}\left[\overline{Y}^2\right]+\frac{1}{2}\left(\lambda-\mathbb{E}\left[T\right]\right)^2 \nonumber \\
&+\left(\lambda-\mathbb{E}\left[T\right]\right)\mathbb{E}\left[T\right]+\frac{1}{2}\mathbb{E}\left[T^2\right] \nonumber \\
=&\frac{1}{2}\lambda^2+\frac{1}{2}\mathbb{E}\left[T^2\right]-\frac{1}{2}\left(\mathbb{E}\left[T\right]\right)^2-\frac{1}{2}\mathbb{E}\left[\overline{Y}^2\right]. \label{eq_thm_exp_Q_1}
\end{align}
The above can be further simplified by noting that using (\ref{eq_exp_x}) and (\ref{eq_exp_x2}) we have
\begin{align}
\mathbb{E}&\left[T^2\right]-\left(\mathbb{E}\left[T\right]\right)^2 \nonumber \\
=&\left(\frac{2}{p}-1\right)\left(\frac{1}{p}-1\right)\gamma^2+2\left(\frac{1}{p}-1\right)\gamma\mathbb{E}\left[\overline{Y}\right]+\mathbb{E}\left[\overline{Y}^2\right] \nonumber \\
&-\left(\frac{1}{p}-1\right)^2\gamma^2-2\left(\frac{1}{p}-1\right)\gamma\mathbb{E}\left[\overline{Y}\right]-\left(\mathbb{E}\left[\overline{Y}\right]\right)^2 \nonumber \\
=&\frac{1}{p}\left(\frac{1}{p}-1\right)\gamma^2+\mathbb{E}\left[\overline{Y}^2\right]-\left(\mathbb{E}\left[\overline{Y}\right]\right)^2 \nonumber \\
=&\frac{1-p}{p^2}\gamma^2+\mathbb{E}\left[\overline{Y}^2\right]-\left(\mathbb{E}\left[\overline{Y}\right]\right)^2,
\end{align}
which, upon substituting in (\ref{eq_thm_exp_Q_1}) finally gives
\begin{align}
\mathbb{E}\left[Q\right]=\frac{1}{2}\lambda^2+\frac{1-p}{2p^2}\gamma^2-\frac{1}{2}\left(\mathbb{E}\left[\overline{Y}\right]\right)^2. \label{eq_thm_exp_Q_2}
\end{align}

Now using (\ref{eq_thm_exp_L}) and (\ref{eq_thm_exp_Q_2}) we have
\begin{align}
g(\lambda)=&\mathbb{E}\left[Q\right]-\lambda\mathbb{E}\left[L\right] \nonumber \\
=&-\frac{1}{2}\lambda^2+\frac{1-p}{2p^2}\gamma^2-\frac{1}{2}\left(\mathbb{E}\left[\overline{Y}\right]\right)^2+\lambda\mathbb{E}\left[\overline{Y}\right].
\end{align}
Thus, solving $g(\lambda^*)=0$ is equivalent to solving
\begin{align}
\left(\lambda^*\right)^2-2\mathbb{E}\left[\overline{Y}\right]\lambda^*+\left(\mathbb{E}\left[\overline{Y}\right]\right)^2-\frac{1-p}{p^2}\gamma^2=0.
\end{align}
The above equation has two solutions, but only one of them is valid due to the inequality in (\ref{eq_thm_lmda_lb}). This is given by
\begin{align}
\lambda^*=\mathbb{E}\left[\overline{Y}\right]+\frac{\sqrt{1-p}}{p}\gamma.
\end{align}

It now remains to check whether $\gamma<\lambda^*-\mathbb{E}\left[T\right]$ holds. Using (\ref{eq_exp_x}), we have
\begin{align}
\lambda^*-\mathbb{E}\left[T\right]=\frac{\sqrt{1-p}}{p}\gamma-\frac{1-p}{p}\gamma,
\end{align}
which is clearly no larger than $\gamma$ since the quantity $\frac{\sqrt{1-p}-\left(1-p\right)}{p}$ is no larger than $1$ for all values of $p$. This gives a contradiction and concludes the proof.

\subsection{Different Methods for Deriving (\ref{eq_w_eta})} \label{apndx_w_eta}

The method included in the main text to derive (\ref{eq_w_eta}) involves a calculus of variations approach mainly through leveraging the Euler-Lagrange equation and equating the functional derivative to $0$ \cite{luenberger-optimization}. In this appendix we discuss two alternate methods to derive (\ref{eq_w_eta}).

The first method, and quite the simplest one, is by completing the square in the Lagrangian in (\ref{eq_lagrangian}). Specifically, (\ref{eq_lagrangian}) can be rewritten equivalently as
\begin{align}
\mathcal{L}=&\int_c^\gamma\frac{1}{2}\left(w(\tau)+\tau+\mathbb{E}\left[T\right]-\lambda-\frac{\eta(\tau)}{f_Y(\tau)}\right)^2f_Y(\tau)d\tau \nonumber \\
&-\int_c^\gamma\frac{1}{2}\left(\tau+\mathbb{E}\left[T\right]-\lambda-\frac{\eta(\tau)}{f_Y(\tau)}\right)^2f_Y(\tau)d\tau \nonumber \\
&+\mathbb{E}\left[\overline{Y}\right]\mathbb{E}\left[T\right]+\frac{1}{2}\mathbb{E}\left[T^2\right]-\lambda\mathbb{E}\left[T\right],
\end{align}
which is minimized iff the first integrand is set to $0$ $\forall \tau$, which exactly gives (\ref{eq_w_eta}).

The second method is by using the result in \cite[Ch.~7 Th.~1]{luenberger-optimization} to conclude that $\mathcal{L}\left(w\right)$ is minimized at $w^*$ only if
\begin{align} \label{eq_lagrangian_Gateaux}
\left.\frac{\partial}{\partial \alpha}\mathcal{L}\left(w^*+\alpha h\right)\right|_{\alpha=0}=0,
\end{align}
for any $h(\cdot):\mathbb{R}_+\rightarrow\mathbb{R}_+$. Taking $h(\tau)\triangleq\delta(\tau-t)$, for some $t\in[c,\gamma]$, where $\delta(\cdot)$ is the Dirac delta function, we get that for fixed $\alpha$
\begin{align} 
\mathcal{L}\!\left(w+\alpha h\right)\!=&\!\!\int_c^\gamma\!\!\left(\left(\tau+\mathbb{E}\left[T\right]-\lambda\right)w(\tau)+\frac{1}{2}w^2(\tau)\right)f_Y(\tau)d\tau \nonumber \\
&+\alpha\left(t+\mathbb{E}\left[T\right]-\lambda\right)f_Y(t)\int_c^\gamma\delta(\tau-t)d\tau \nonumber \\
&+\frac{1}{2}\int_c^\gamma\left(w(\tau)+\alpha\delta(\tau-t)\right)^2f_Y(\tau)d\tau \nonumber \\
&+\mathbb{E}\left[\overline{Y}\right]\mathbb{E}\left[T\right]+\frac{1}{2}\mathbb{E}\left[T^2\right]-\lambda\mathbb{E}\left[T\right] \nonumber \\
&-\int_c^\gamma w(\tau)\eta(\tau)d\tau-\alpha\eta(t)\int_c^\gamma \delta(\tau-t)d\tau.
\end{align}
Therefore, upon using $\int_c^\gamma\delta(\tau-t)d\tau=1$, we have
\begin{align}
\frac{\partial\mathcal{L}\left(w+\alpha h\right)}{\partial\alpha}=&\left(t+\mathbb{E}\left[T\right]-\lambda\right)f_Y(t)-\eta(t) \nonumber \\
&+\int_c^\gamma\left(w(\tau)+\alpha\delta(\tau-t)\right)\delta(\tau-t)f_Y(\tau)d\tau.
\end{align}
Setting $\alpha=0$ in the above and using (\ref{eq_lagrangian_Gateaux}), (\ref{eq_w_eta}) is directly reached after rearranging.


\end{document}